\documentclass[twocolumn]{aa} 
\usepackage{graphicx}
\usepackage{slashbox}
\usepackage{amsmath}
\usepackage{color}
\usepackage{mathptmx} 
\usepackage{multirow}
\usepackage{wasysym}
\usepackage{url}
\usepackage{txfonts}
\usepackage{natbib}

   \title{Asteroid flux towards circumprimary habitable zones in binary star systems: II. Dynamics}
   \author{D. Bancelin
          \inst{1,2}
          \and E. Pilat-Lohinger\inst{1}
          \and \'A. Bazs\'o\inst{1}
          }

   \institute{Institute of Astrophysics (ifA), University of Vienna, T\"urkenschanzstr. 17, A-1180 Vienna,
Austria (\email{david.bancelin@univie.ac.at})
 \and
 IMCCE, Paris Observatory, UPMC, CNRS, UMR8028, 77, Av. Denfert-Rochereau F-75014 Paris,
France
\\
             }

   \date{Received ---, 2011; accepted ---}
   
\def\apjl{ApJ}
\def\apjs{ApJS}
\def\ao{Appl.~Opt.}
\def\aap{A\&A}

\definecolor{gray}{gray}{0.62}
\begin{document}

%


  \abstract
   {Secular and mean motion resonances (MMR) are effective perturbations for shaping planetary systems. In 
binary star systems, they play a key role during the early and late phases of planetary formation, as well as for the 
dynamical stability of a planetary system.}
   {In this study, we aim to correlate the presence of orbital resonances with the rate of icy asteroids crossing the 
habitable zone (HZ) from a circumprimary disk of planetesimals in various binary star systems.}
   {We modelled a belt of small bodies in the inner and outer regions, interior and exterior to the orbit of a gas giant 
planet, respectively. The planetesimals are equally placed around a primary G-type star and move under the gravitational 
influence of the two stars and the gas giant. We numerically integrated the system for 50 Myr, considering various 
parameters for the secondary star. Its stellar type varies from a M- to F-type; its semimajor axis is either 50 au or 
100 au, and its eccentricity is either 0.1 or 0.3. For comparison, we also varied the gas giant's orbital and physical 
parameters.}
{Our simulations highlight that a disk of planetesimals will suffer from perturbations owing to a perturbed gas giant, 
mean motion, and secular resonances. We show that a secular resonance  -- with location and width varying according to 
the secondary star's characteristics -- can exist in the icy asteroid belt region and overlap with MMRs, which  have 
an impact on the dynamical lifetime of the disk. In addition, we point out that, in any case, the 2:1 MMR, the 5:3 MMR, 
and the secular resonance are powerful perturbations for the flux of icy asteroids towards the HZ and the transport of 
water therein.}
   {}
   \titlerunning{Asteroids flux to HZ in binary star systems} 
   \maketitle

\section{Introduction}

Orbital resonances (mean motion and secular resonances) play a key role in the architecture of a planetary 
system. It is well known that they had a strong influence on the dynamics in the early stage \citep{walsh11} 
and late stage \citep{tsiganis05,gomes05} of the planetary formation in our solar system. More generally, the late 
phase 
of planetary formation around single stars is mainly dominated by mean motion resonances (MMRs) as a gas giant 
could have formed within 10 
Myr \citep{briceno01}. \\

The role of a secular perturbation is mainly critical during the early phases of 
planetary formation in binary systems \citep[and references therein]{thebault14} since it influences the collisional 
velocities between planetesimals. However, if planets manage to form despite these strong perturbations, their 
dynamical outcome will anyway be governed by MMR and secular resonances, which will determine the well-known stability 
criteria in binary star systems \citep{rabl88,holman99,pilat02,mudryk06}. 
The location of orbital 
resonances have to be known to predict the fate of bodies evolving nearby or inside. The 
location of the secular resonance can be determined analytically by the Laplace-Lagrange perturbation 
theory \citep[e.g.][]{murray99} provided that the planets
have low eccentricities and inclinations and negligible masses compared to the central star. In this context, 
\cite{pilat08} 
analyzed the effect of different Jupiter-Saturn configurations by varying the mutual distance and the mass-ratio of 
the two planets. They 
showed that, for some configurations, the habitable zone (HZ) of the Sun 
would be affected by a 
secular resonance whose frequency $g = g_{\scriptscriptstyle 
{\text{Jupiter}}}$, with $g_{\scriptscriptstyle 
{\text{Jupiter}}}$ the proper secular frequency of Jupiter (as defined in the Laplace-Lagrange theory). Moreover, it 
was shown 
that the analytical result was in  
good agreement with that of the numerical study. An application of this secular 
perturbation theory to circumstellar planetary motion in binary star systems 
causes problems in the accuracy owing to the massive secondary star, which often 
moves in an eccentric orbit around the center of mass. \cite{pilat15} show that the secular resonance that occurs in tight binary star 
systems hosting a giant planet \citep[see e.g.][]{pilat05} can be located by a 
semi-analytical method. This method uses the Laplace-Lagrange perturbation 
theory to determine the proper frequencies of test planets in these binary 
star--planet configurations, while the frequency of the giant planet is 
calculated from a numerical time series via a fast Fourier transformation. The 
location of the secular resonance is given by the intersection of the two 
resulting curves. An application of this method to all known binary star systems 
hosting a giant planet in circumstellar motion is presented in the paper by 
\cite{bazso15}.\\

Icy bodies trapped into orbital resonances could be potential water sources for planets in the HZ. 
These water rich objects can be embryos \citep{haghighipour07} and small bodies (asteroids) as shown in 
\cite{morbidelli00} and \cite{obrien14} for our Earth and more recently, in \cite{bancelin15} (hereafter Paper 
I), in binary star systems. In 
this latter study, the author gives a statistical estimate of the contribution of asteroids in bearing water material to 
the HZ. Previous studies in binary 
star systems \citep[e.g.][]{haghighipour07} focused mainly on the water transport via embryos.\\

This study is a continuation of the investigation from Paper I and elucidates dynamical features of 
this work. In Paper I, we showed a statistical overview about the flux of asteroids towards the circumprimary HZ of a 
binary star systems. Considering different binary configurations, we determined the timescale for an asteroid to reach 
the HZ (which is in the range between a few centuries and some ten thousands years) and provided estimates for the 
probability for an asteroid to cross the HZ (1 -- 50\%) and the quantity of water brought to the HZ 
(varying by a factor of 1 to 15 depending on the binary set-up). 
In the present study, to explain the differences in the statistical results, we aim to emphasize and characterize the dynamical effects of 
orbital resonances on a 
disk of 
planetesimals, in various binary star systems hosting a gas giant planet, as well as to what extent such 
resonances are likely to enable icy asteroids to bring water material into the 
HZ in comparison to single star systems. In Section \ref{S:disk}, we define our initial 
modelling for the binary 
star systems, the gas giant, and the disk of planetesimals. Then in Section \ref{S:dynamics}, we discuss their 
dynamical outcome. We also analyze the dynamical behaviour of particles initially orbiting near or 
inside MMRs to highlight the discrepancies on the dynamical lifetime according to the binary star system investigated. 
In Section \ref{S:flux}, we combine our results to analyze the consequences of such dynamics on the flux of 
icy asteroids towards the HZ and the water transport therein. We then analyze how the previous results are 
influenced when changing the orbital and physical parameters of the gas giant (Section \ref{S:Jupiter}) or when one or 
two giant planets orbit around one G2V star (section \ref{S:single}). Finally, we provide a comparison of the 
water transport efficiency between binary and single star systems in Section \ref{S:comparison} and conclude 
our work in Section \ref{S:conclusion}.

\section{Dynamical model}\label{S:disk}

We focused our work on a primary G-type star but the secondary is either an F-, G-, K- or M-type star with mass 
$M_{\scriptscriptstyle \text{b}}$ equal to $1.3M_{\scriptscriptstyle \odot}$, $1.0M_{\scriptscriptstyle 
\odot}$, $0.7M_{\scriptscriptstyle \odot}$ and $0.4M_{\scriptscriptstyle \odot}$, respectively. The orbital separations 
are 
$a_{\scriptscriptstyle \text{b}}$ = 50 au and 100 au. The secondary's eccentricity is $e_{\scriptscriptstyle \text{b}}$ 
= 0.1 and 0.3; its inclination is set to 0$^{\circ}$. Our systems also host  a gas giant planet initially at 
$a_{\scriptscriptstyle {\text {GG}}}$ = 5.2 au moving on a circular orbit, in the same plane as the secondary, and with 
a mass $M_{\scriptscriptstyle \text{GG}}$ = 1$M_{\scriptscriptstyle \text{J}}$, $M_{\scriptscriptstyle 
\text{J}}$ being the mass of Jupiter. \\

We modelled a disk of planetesimals inside and beyond the orbit of the gas giant. To avoid strong initial 
interaction with the gas giant, we assumed that it has gravitationally cleared a path in the disk around its orbit. We 
defined the width 
of this path as $\pm 3\,R_{\scriptscriptstyle {\text{H},\text{GG}}}$, where $R_{\scriptscriptstyle 
{\text{H},\text{GG}}}$ is the giant planet's Hill radius. Contrary to Paper I, in which the small bodies were 
randomly placed beyond the snow line, we considered a different initial set up of the planetesimal disk to allow easy comparisons of the 
dynamics in the various systems. We defined three different regions in our planetesimal disk:
\begin{itemize}
 \item $\mathcal{R}_{1}$: this region extends from 0.5 au to the snow line position 
at $\sim$ 2.7 au (value for a primary G-type star). 200 particles were initially placed in this region.
 \item $\mathcal{R}_2$: this region extends from beyond the snow line and up to the distance $a_{\scriptscriptstyle 
{\text {GG}}}- 3\,R_{\scriptscriptstyle 
{\text{H},\text{GG}}}$ $\approx$ 4.1 au. We define this region as the inner disk. As we are mainly interested in icy 
bodies that are likely to bring water to the HZ, we densified this region and 1 000 particles were distributed therein.
 \item $\mathcal{R}_3$: this region extends from $a_{\scriptscriptstyle {\text {GG}}} + 3\,R_{\scriptscriptstyle 
{\text{H},\text{GG}}}$ $\approx$ 6.3 au and up to the 
stability 
limit criteria defined by the critical semimajor axis $a_{\scriptscriptstyle {\text {c}}}$ \citep{holman99,pilat02} and 
 its variance $\Delta\,a_{\scriptscriptstyle {\text {c}}}$. For the studied 
systems, the value of $a_{\scriptscriptstyle {\text {c}}}$ are given in Table 
\ref{T:ac} using the method of \cite{pilat02}. It is obvious that the size of 
the external disk will vary according to ($a_{\scriptscriptstyle {\text {b}}}$, $e_{\scriptscriptstyle {\text {b}}}$, 
$M_{\scriptscriptstyle {\text {b}}}$). The larger $a_{\scriptscriptstyle {\text {b}}}$ and the smaller 
($e_{\scriptscriptstyle {\text{b}}}$, $M_{\scriptscriptstyle {\text {b}}}$), 
the wider this region. We 
defined this region as the outer disk in which 1 000 particles were placed.
\end{itemize}
\begin{table}
 \begin{center}
  \caption{Stability limits $a_{\scriptscriptstyle \text{c}}$ for the 
considered binary configurations. For the values of $a_{\scriptscriptstyle 
\text{c}}$, we applied the results by \cite{pilat02}}
  \label{T:ac}
  \begin{tabular}{cccc}
   \hline
   \hline
Secondary star & $e_{\scriptscriptstyle \text{b}}$  & $a_{\scriptscriptstyle 
\text{c}}$ [au] & $a_{\scriptscriptstyle \text{c}}$ [au] \\
               &                                    & ($a_{\scriptscriptstyle 
\text{b}}$ = 50 au) & ($a_{\scriptscriptstyle 
\text{b}}$ = 100 au)\\
\hline
 F & 0.1 & 11.9 & 23.8 \cr
   & 0.3 & 7.9  & 11.8 \cr
\hline
 G & 0.1 & 12.5 & 25.0 \cr
   & 0.3 & 8.0  & 16.0 \cr
\hline 
 K & 0.1 & 13.4 & 26.8\cr
   & 0.3 & 8.9 & 17.8 \cr
 \hline
 M & 0.1 & 14.3 & 28.6 \cr
   & 0.3 & 10.3 & 20.7 \cr

\hline
  \end{tabular}
 \end{center}
\end{table}

For all three cases, the initial orbital separation between each particle is uniform and is defined as the ratio 
between the width of each region and the number of particles. Their initial motion is taken as nearly circular 
and planar. We also assumed all our asteroids in $\mathcal{R}_2$ and $\mathcal{R}_3,$ with equal mass and 
an initial water mass fraction of 10\%. Water mass-loss owing to ice sublimation was also taken into account during the 
numerical integrations. All 
the 
simulations are purely gravitational since we consider our numerical integrations to start as soon as the gas 
vanished (therefore, we do not consider gas driven migration and eccentricity dampening). We also assumed that, at this 
stage, planetary embryos have been able to form. Our simulations were performed for 50 Myr using 
the Radau integrator in the Mercury6 package \citep{chambers99}.

\section{Dynamics under the binary star perturbation}\label{S:dynamics}

\subsection{Orbit of the giant planet}

\begin{figure}[h!]
 \centering{\includegraphics[angle=-90,width=\columnwidth]{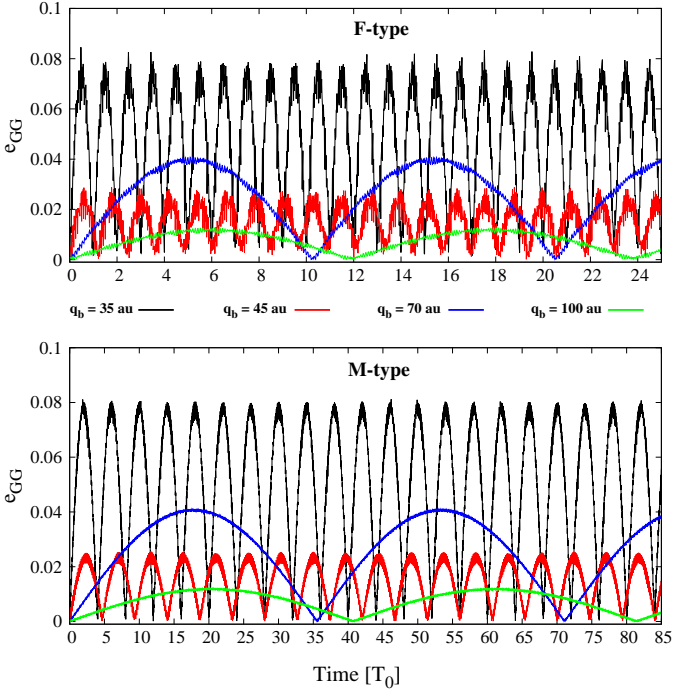}}
  \caption{Variation of the gas giant's eccentricity $e_{\scriptscriptstyle {\text{GG}}}$, under the perturbation of a 
secondary F-star (top) and M-star (bottom), as a function of time (expressed in 
$T_{\scriptscriptstyle 0}$ units, see text) and for different secondary's periapsis distance $q_{\scriptscriptstyle 
{\text{b}}}$.} 
    \label{F:ecc_giant}
\end{figure}

\begin{figure}[h!]
 \centering{\includegraphics[angle=-90,width=0.75\columnwidth]{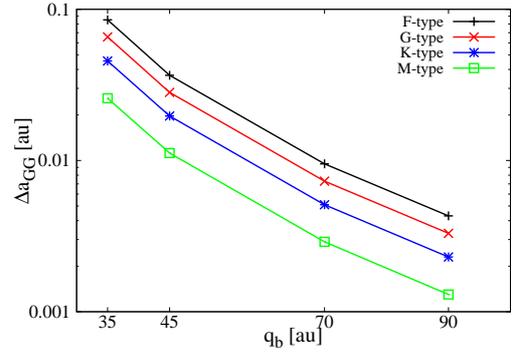}}
  \caption{Orbital inward shift $\Delta\,a_{\scriptscriptstyle \text{GG}}$ of the gas giant as a function of the 
secondary's periapsis $q_{\scriptscriptstyle \text{b}}$ and stellar type.}
     \label{F:giant}
\end{figure}

Owing to the secondary star, it is obvious that the gas giant will no longer remain on a circular motion. This is 
illustrated
in Fig. \ref{F:ecc_giant}, where the variation of the gas giant's eccentricity $e_{\scriptscriptstyle \text{GG}}$ is 
plotted for different values of 
the secondary's periapsis distance $q_{\scriptscriptstyle \text{b}}$ (35 au, 45 au, 70 au, and 90 au) and for two 
different masses $M_{\scriptscriptstyle \text{b}}$ 
(F- and M-type, on the top and bottom panel, respectively). According to the secular theory, the 
variations of e$_{\scriptscriptstyle {GG}}$ are linked to the oscillation period $T$ of the giant planet and the
amplitude depends on the mass and 
periapsis distance of the perturber. The higher $M_{\scriptscriptstyle \text{b}}$ and the smaller 
$q_{\scriptscriptstyle 
\text{b}}$, the shorter $T$. In 
our case, $T$ will be shorter for the largest secondary's mass investigated (F-type) and 
smallest value of $q_{\scriptscriptstyle \text{b}}$ (35 au). Therefore, we scaled the x-axis to this 
period\footnote{The 
numerical value is $T_{\scriptscriptstyle 0}$ = 6323 yr} $T_{\scriptscriptstyle 0}$. For a given value of 
$q_{\scriptscriptstyle {\text{b}}}$, we can see that the secondary's mass does not have a major 
impact on the maximum value\footnote{The difference does not exceed $\sim\,0.5\%$} of $e_{\scriptscriptstyle 
{\text{GG}}}$. For instance, for $q_{\scriptscriptstyle {\text{b}}}$ = 45 au, we found that $e_{\scriptscriptstyle 
{\text{GG}}} \sim 0.04$ for any value of $M_{\scriptscriptstyle {\text{b}}}$. 
However, its period 
increases with 
$q_{\scriptscriptstyle {\text{b}}}$ and $M_{\scriptscriptstyle {\text{b}}}$. Indeed, for $q_{\scriptscriptstyle 
{\text{b}}}$ = 35 au, 
$T= T_{\scriptscriptstyle 0}$ for a secondary F-type and $T=4\,T_{\scriptscriptstyle 0}$ if the secondary is a 
M-type star. Also, for $q_{\scriptscriptstyle {\text{b}}}$ = 90 au, we have 
$T= 12\,T_{\scriptscriptstyle 0}$ and $T=40\,T_{\scriptscriptstyle 0}$ for a secondary F- and M-type star, 
respectively. 
These periodic 
variations of $e_{\scriptscriptstyle {\text{GG}}}$ will have consequences on the gravitational interaction with 
the disk of planetesimals.\\
The interaction with the disk will be strengthened since the gas giant can encounter a semimajor axis drift. Indeed, we 
show in Fig. \ref{F:giant} that the dynamical perturbations induced by the secondary can slightly shift  the 
initial semimajor axis of the gas giant inward by a quantity $\Delta\,a_{\scriptscriptstyle {\text{GG}}}$. Therefore, the 
location of MMRs will be affected since they mainly depend on $a_{\scriptscriptstyle {\text{GG}}}$. The consequences of 
this type of drift will be analyzed in Section \ref{S:MMR}.
\subsection{Dynamics of the disk of planetesimals}

\begin{figure*}
\centering{
    \begin{tabular}{cc}
      \includegraphics[angle=-90,width=0.45\textwidth]{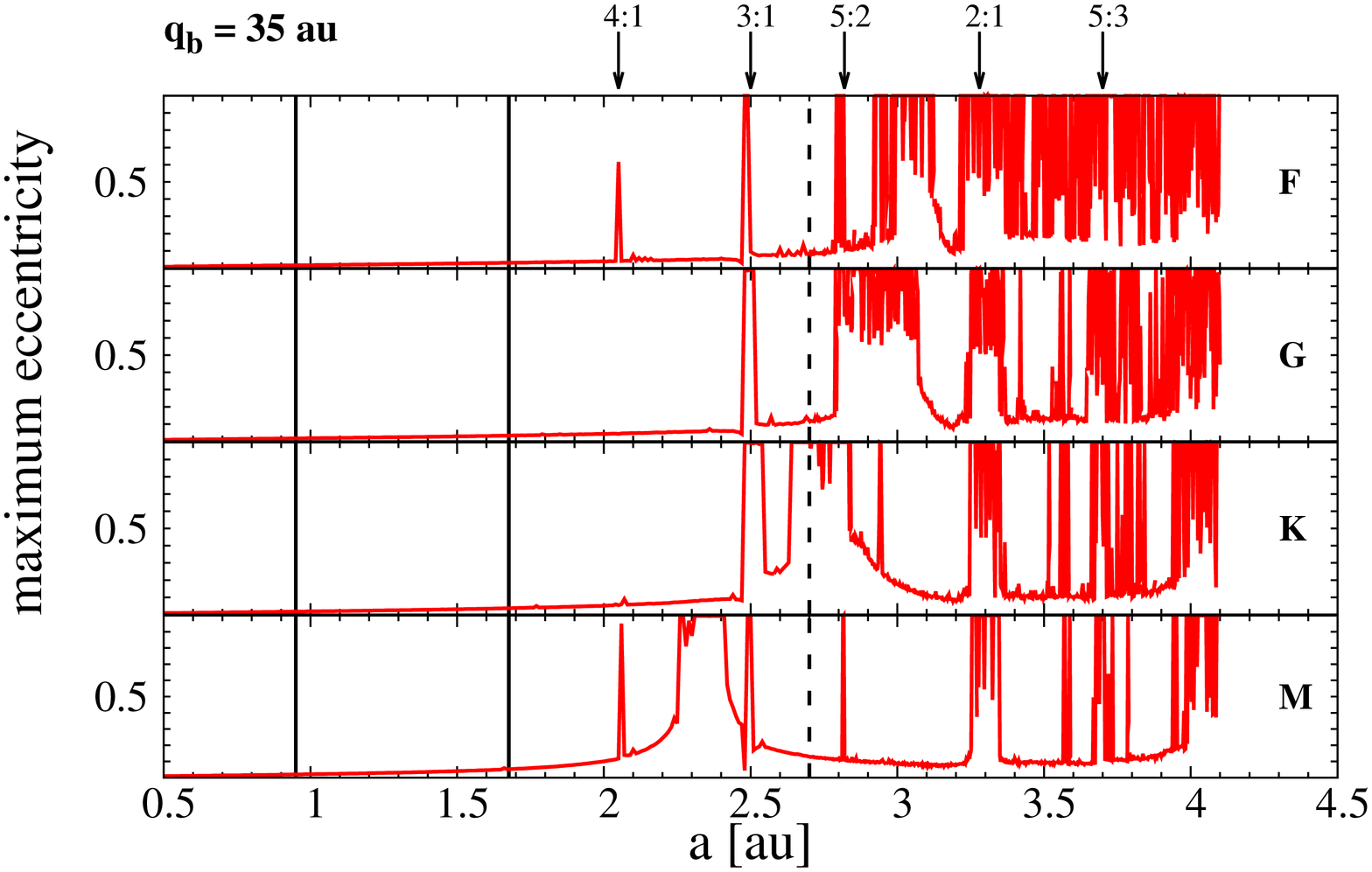} &
      \includegraphics[angle=-90,width=0.45\textwidth]{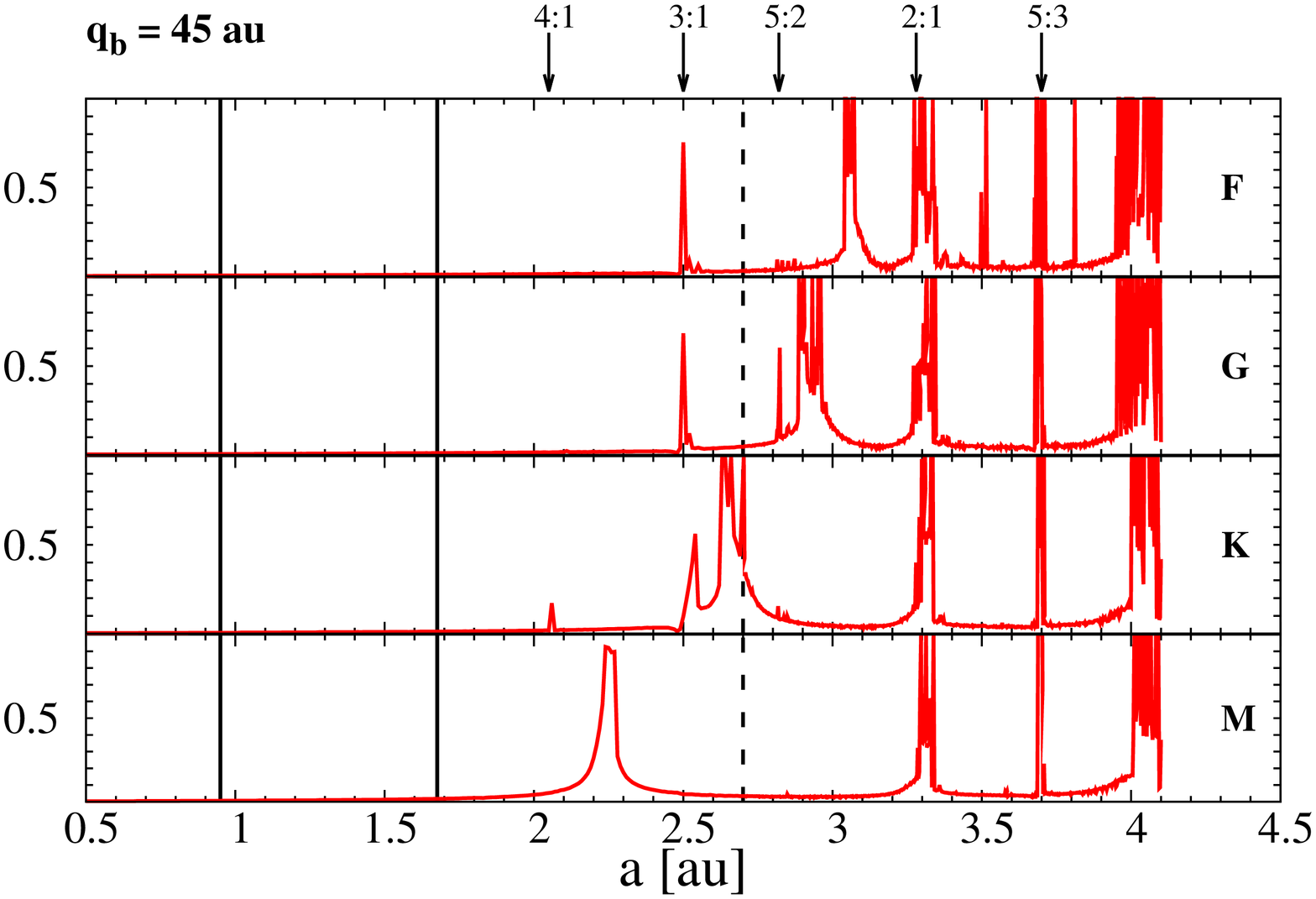} \\
      \includegraphics[angle=-90,width=0.45\textwidth]{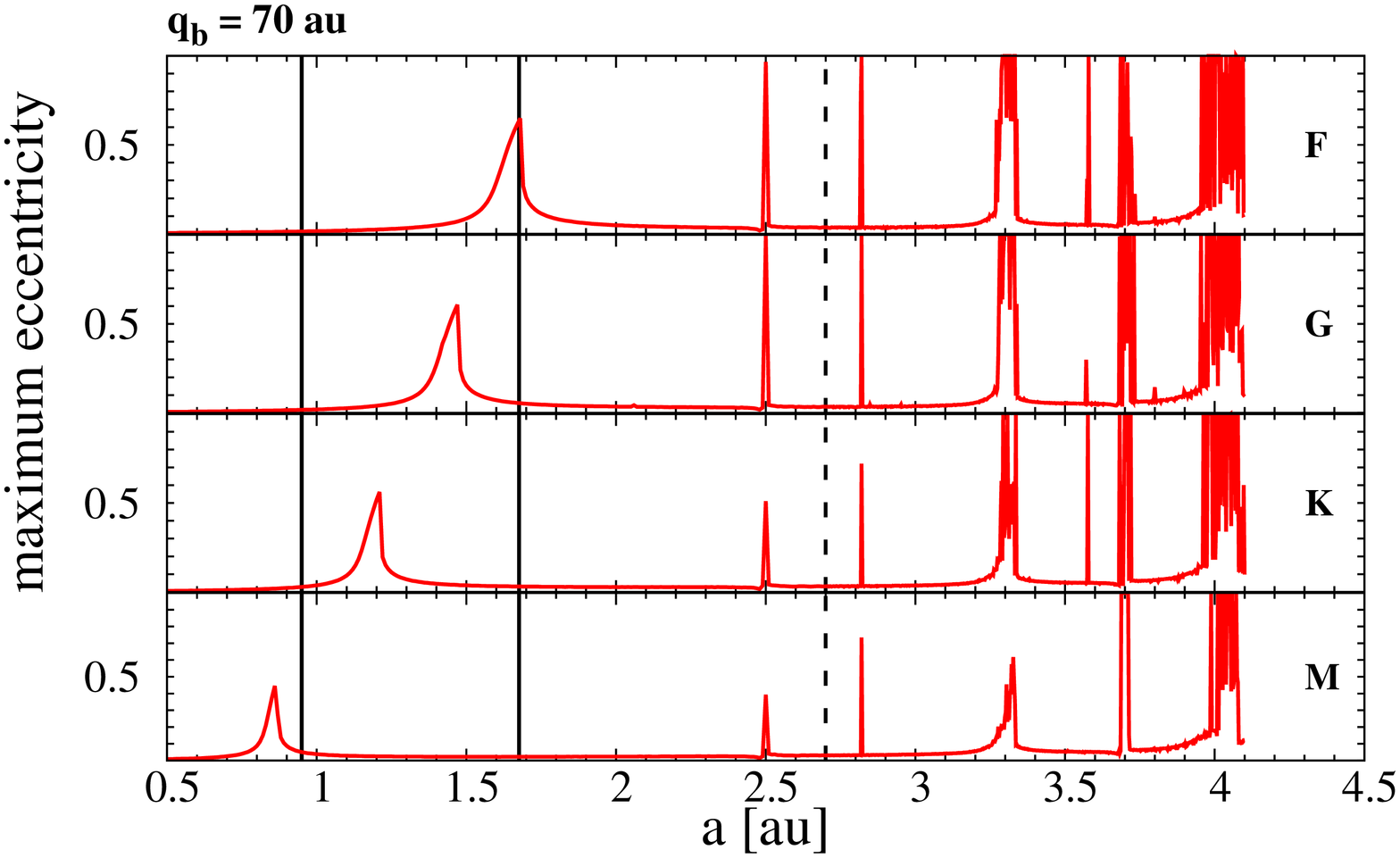} &
      \includegraphics[angle=-90,width=0.45\textwidth]{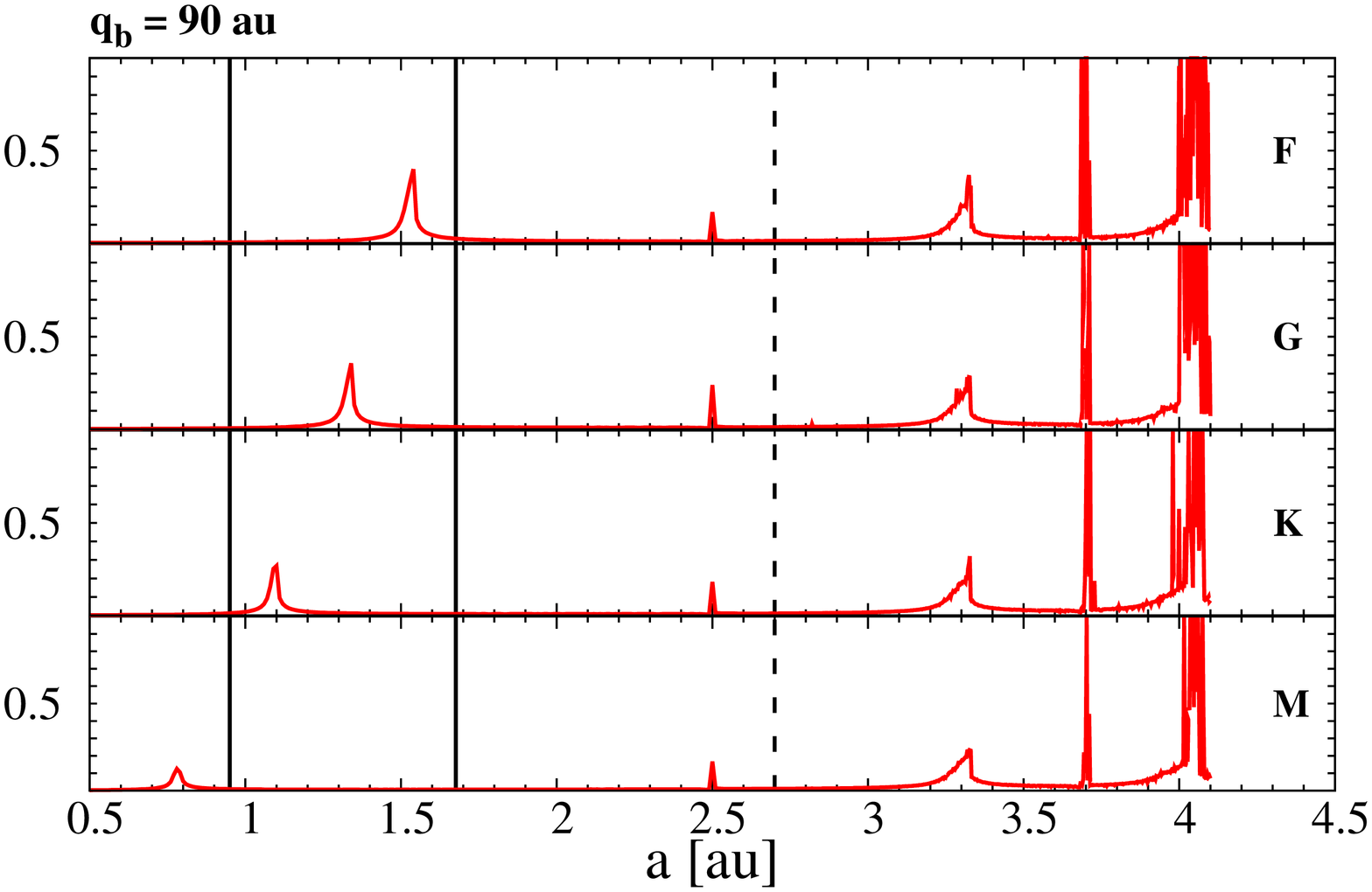} 
    \end{tabular}}
    \caption{Maximum eccentricity of planetesimals in the $\mathcal{R}_1$ and $\mathcal{R}_2$ regions, separated by the 
dashed vertical line for the snow-line position, as a function of 
their initial position, up to an intermediate integration time of 5 Myr, for different values of 
$q_{\scriptscriptstyle {\text{b}}}$. Each subpanel refers to the 
secondary stellar type and the continuous vertical lines refer to the HZ borders. In addition, the inner main MMRs 
with 
the gas giant are indicated.}\label{F:e_inner}
\end{figure*}

In Fig. \ref{F:e_inner}, we show the maximum eccentricity reached by the planetesimals at different 
initial semimajor axes, in the regions $\mathcal{R}_1$ and $\mathcal{R}_2$ (separated by the dashed vertical line 
representing the snow-line position). The four panels correspond to the values of 
$q_{\scriptscriptstyle \text{b}}$ investigated and each subpanel is for 
different secondary stellar types (F, G, K, and M). We can distinguish 
MMRs\footnote{Only the main ones are 
indicated} with the gas giant 
and also a secular resonance: on the bottom panels, which represent the results 
for $a_{\scriptscriptstyle \text{b}}$ = 100 au ($q_{\scriptscriptstyle \text{b}}$ = 70 au and 
$q_{\scriptscriptstyle \text{b}}$ = 90 au), we can see a spike located close to or inside the HZ\footnote{The 
borders are defined according to \cite{kopparapu13}}(continuous vertical lines) and moving outward (to larger 
semi-major 
axes) when increasing the secondary's mass. This spike represents the secular resonance. When increasing 
$q_{\scriptscriptstyle \text{b}}$, not only does it slightly move 
inward 
but also, the maximum eccentricity reached by the particles is higher. This is because the gravitational perturbation 
from the 
secondary increases the gas giant's eccentricity. As a consequence, the forced eccentricity 
contribution of 
any particle inside the secular resonance will be increased. When decreasing $a_{\scriptscriptstyle \text{b}}$ to 50 au 
(top panels for 
$q_{\scriptscriptstyle \text{b}}$ = 35 au and $q_{\scriptscriptstyle \text{b}}$ = 45 au), the secular resonance 
moves also 
outward 
and reaches the MMR region. As a consequence, the inner disk will suffer from an overlap of these orbital resonances 
that could cause a fast depletion.  However, particles inside the HZ will remain on near circular motion. We should 
also notice, as studied in detail by \cite{pilat15}, the combined effects 
induced by a change of $e_{\scriptscriptstyle 
\text{b}}$ when the other dynamical parameters ($a_{\scriptscriptstyle \text{b}}$ and $M_{\scriptscriptstyle 
\text{b}}$) remain constant. Indeed, an increase of $e_{\scriptscriptstyle 
\text{b}}$ will turn into: 
\begin{itemize}
 \item an increase of $e_{\scriptscriptstyle {\text{GG}}}$ (see Fig. 
\ref{F:ecc_giant}) so that the size of the secular 
perturbed area 
is increased but the location remains the same;
 \item an inward shift of the gas giant with intensity depending on the binary star characteristics (see Fig. 
\ref{F:giant}) where the width of the secular resonance remains unchanged 
but its location is shifted inward.
\end{itemize}
As a consequence, only the width (see top panels in Fig. \ref{F:e_inner}) or both the location and the width 
of the secular resonance can be modified (see bottom panels in Fig. \ref{F:e_inner}), as also shown in 
\cite{pilat15} and \cite{bazso15}.\\
The outer disk exhibits dynamical outcomes different from the inner disk. 
Indeed, only external 
MMRs\footnote{MMRs with the secondary star also 
influence the dynamics of celestial bodies and their location depends on the secondary star's orbital parameters. 
However, they have a very high order and their contribution is much weaker compared to the MMRs with the gas giant.} 
and gravitational interactions  with the gas giant perturb $\mathcal{R}_3$. 
As shown in Fig. \ref{F:ecc_giant}, in the case of $q_{\scriptscriptstyle 
{\text{b}}}$ = 35 au, the periodic 
high variations of $e_{\scriptscriptstyle {\text{GG}}}$, coupled with the gravitational excitation from the 
secondary, will cause a strong interaction with the external disk and its dynamical behaviour will be more or less 
chaotic. When increasing $q_{\scriptscriptstyle {\text{b}}}$ (and therefore increasing the size of the disk and 
decreasing $e_{\scriptscriptstyle {\text{GG}}}$), the external MMRs dominate.\\
\begin{figure*}
\centering{\includegraphics[angle=-90,width=0.75\textwidth]{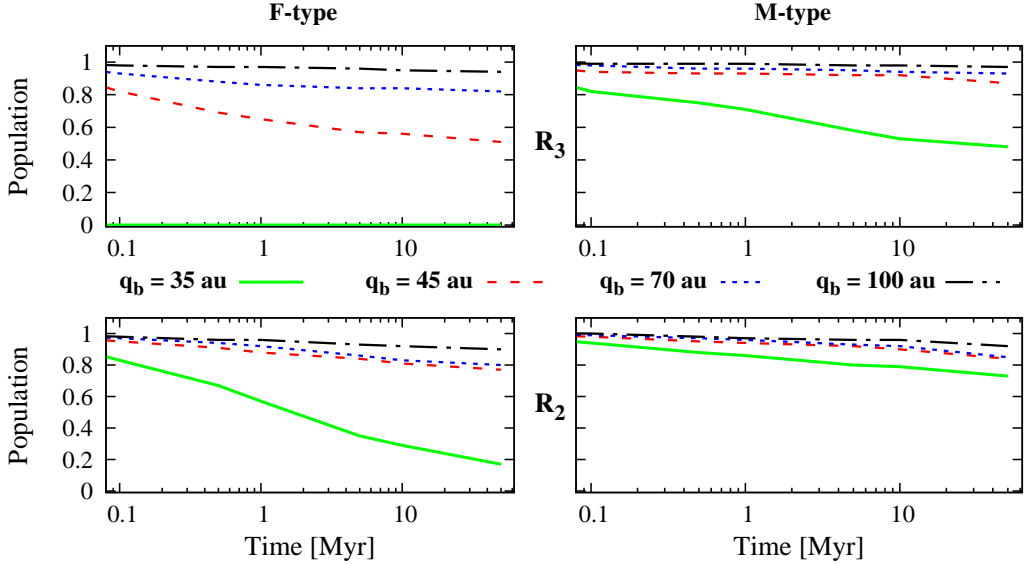}}
  \caption{Evolution of the remaining population in the $\mathcal{R}_3$(top) and $\mathcal{R}_2$(bottom) 
regions according to the periapsis 
distance $q_{\scriptscriptstyle {\text{b}}}$ of a secondary F-type (left) and M-type (right) star.}
     \label{F:pop}
\end{figure*}
To highlight our results, in Fig. \ref{F:pop}, we show the dynamical outcome of the $\mathcal{R}_2$ and $\mathcal{R}_3$ 
regions for a secondary F-type (left panel) and M-type (right panel) star. 
Because of the definition of 
$\mathcal{R}_3$, a secondary F-type star at $q_{\scriptscriptstyle {\text{b}}}$ 
= 35 au cannot host an outer 
disk\footnote{as $a_{\scriptscriptstyle {\text{c}}}$ - $\Delta$ $a_{\scriptscriptstyle {\text{c}}}$ $\le$ 
$a_{\scriptscriptstyle {\text {GG}}} + 3\,R_{\scriptscriptstyle {\text{H},\text{GG}}}$}. For this 
value of $q_{\scriptscriptstyle {\text{b}}}$, we can see that more than 80\% of the inner disk escaped 
within 50 Myr. This is because the secular resonance lies inside $\mathcal{R}_2$ and its large width (see top right 
panel in Fig. \ref{F:e_inner} for the case of a secondary F-type) will favour chaos therein. Thus, this orbital 
resonance 
plays an important role for the depletion of a planetesimal disk. We expect 
similar results for a G- and K-type secondary as the secular resonance lies 
beyond the snow line. This is not the 
case for an M-type at the same periapsis distance and we can clearly see that the lack of strong perturbations 
inside $\mathcal{R}_2$ (the secular resonance is inside the snow line, see top right panel in Fig. 
\ref{F:e_inner} for the case of a secondary M-type) enables 70\% of the inner disk population to survive within 
50 Myr. The evolution of the population in 
$\mathcal{R}_3$ will be strongly correlated to this region's width and to the gravitational 
interactions with the gas giant and 
the secondary star. The more compact  the outer disk is and the stronger  the perturbations are from the massive bodies, 
the more likely the loss of asteroids in $\mathcal{R}_3$.\\

This result can also put in question the presence and observational evidence of a 
remaining asteroid belt in these types of systems for 
$a_{\scriptscriptstyle \text{b}}$ = 50 au, since an inner disk can suffer from both secular and mean motion perturbations, 
but 
also from strong interactions with the gas giant. It might not survive under such conditions since we can observe a linear 
decrease of the population of $\mathcal{R}_2$. Therefore, its dynamical lifetime will vary according to 
$M_{\scriptscriptstyle 
\text{b}}$.

\subsection{Dynamical lifetime of particles near MMRs}\label{S:MMR}

\begin{figure}[h!]
 \centering{\includegraphics[angle=-90,width=\columnwidth]{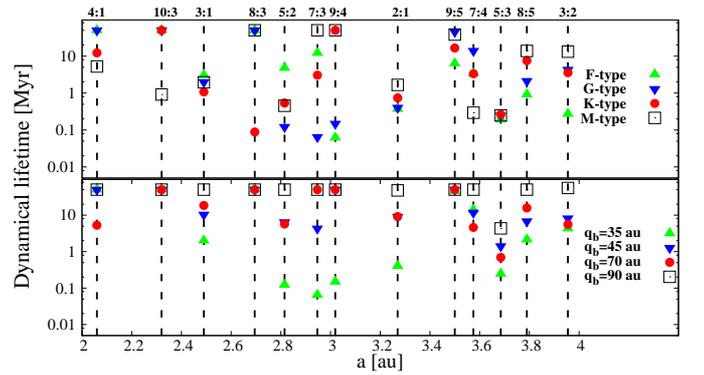}}
  \caption{Dynamical lifetime of test particles in $\mathcal{R}_1$ and $\mathcal{R}_2$ regions expressed in Myr. Top 
panel: influence of the 
secondary's mass when $q_{\scriptscriptstyle {\text{b}}}$ = 35 au. Bottom panel: 
influence of the secondary's periapsis distance $q_{\scriptscriptstyle \text{b}}$  assuming the secondary as a G-type 
star.}
     \label{F:MMR}
\end{figure}

In this section, we investigate in detail the dynamical lifetime of particles which are initially close or inside 
internal MMRs. 
They occur when the orbital periods of the gas giant and the particle are in commensurability, such as
\begin{equation*}
 \displaystyle a _{\scriptscriptstyle \text{n}} = \left (\frac{\text{p}}{\text{q}} \right)^{\scriptscriptstyle 
{2/3}} a_{\scriptscriptstyle {\text{GG}}}\,\mbox{,}~~~\mbox{p and q} \in \mathbb{N} \displaystyle \left \{ 
\begin{array}{r}
                                              \text{p}>\text{q} ~~\mbox{if} ~~\mbox{} a_{\scriptscriptstyle 
\text{n}} < a_{\scriptscriptstyle \text{GG}} \\
                                              \text{p}<\text{q} ~~\mbox{if} ~~\mbox{} a_{\scriptscriptstyle 
\text{n}} > a_{\scriptscriptstyle \text{GG}} 
                                            \end{array} \right.,
\end{equation*}
where $a_{\scriptscriptstyle \text{n}}$ is the position of the nominal resonance 
location. As shown in Fig. \ref{F:giant}, 
the gas giant will have periodic variations in the interval $\left [a_{\scriptscriptstyle 
\text{GG}} - \Delta\,a_{\scriptscriptstyle \text{GG}}; a_{\scriptscriptstyle \text{GG}} \right]$. Thus, 
for a given pair (p,q), the maximum shift $\Delta\,a_{\scriptscriptstyle \text{n}}$ of $a_{\scriptscriptstyle 
\text{n}}$ is $(\text{p}/\text{q})^{\scriptscriptstyle {2/3}} \Delta\,a_{\scriptscriptstyle \text{GG}}$. As we 
aim to correlate the dynamical lifetime of particles with the binary star 
characteristics, we preferred to do a separate analysis to 
ensure that each MMR contains the same number of particles. Indeed, our results would be biased if we were to use 
the sampling described in 
Section \ref{S:disk}. Instead, for each MMR investigated, we defined an 
interval $\cal{A}_{\scriptscriptstyle \text{n}}$ of 
initial semimajor axis for our test particles:
\begin{equation*}
 \displaystyle \mathcal{A}_{\scriptscriptstyle {n}} = \left [ a_{\scriptscriptstyle n} - \Delta\, a_{\scriptscriptstyle 
n} - \epsilon;\, a_{\scriptscriptstyle n} + \epsilon^\prime \right]\,\mbox{,}~~~\mbox{} (\epsilon, \epsilon^\prime) 
\in \mathbb{R,}
\end{equation*}
where $(\epsilon, \epsilon^\prime)$ are arbitrary numbers\footnote{$(\epsilon, \epsilon^\prime) \sim \pm 1\%$} to take into account particles initially orbiting near the MMR and likely to cross it during the integration. We 
limited 
this study to resonances with integers p and q $\le$ 10. In each MMR, we 
uniformly distributed 25 
particles\footnote{We did not put more particles as $\mathcal{A}_{\scriptscriptstyle n}$ is very narrow. In 
$\mathcal{R}_2$, the size of $\mathcal{A}_{\scriptscriptstyle n}$ is $\sim\,\Delta\,a_{\scriptscriptstyle
\text{n}} \le\,\Delta\,a_{\scriptscriptstyle {GG}}$ as $(p/q)^{2/3}\,\le\,1$. In $\mathcal{R}_3$, we would have 
considerered an other approach as the size of $\mathcal{A}_{\scriptscriptstyle n}$ would be much wider because 
$(p/q)^{2/3}\,\ge\,1$} with a step-size that depends on the size of $\mathcal{A}_{\scriptscriptstyle n}$, initially on 
circular and planar orbits. In addition, as suggested by \cite{pilat02}, each particle is cloned into four starting 
points with mean anomalies $M$ = 
0$^\circ$, $M$ = 90$^\circ$, $M$ = 180$^\circ$ and $M$ = 270$^\circ$, since it is 
well known that the starting position plays an important role for the dynamical 
behaviour in MMRs. This accounts for 100 particles in each MMR. Each 
system was integrated for 50 Myr. We did not consider the condition $a \notin \mathcal{A}_{\scriptscriptstyle n}$ as a 
criterion for a particle to leave the MMR. Indeed, its dynamical evolution can be quite random: from time 
to time it can leave an MMR or be temporarily captured into another MMR. Instead, we consider 
a particle as leaving its initial location in a specific MMR when 
its dynamical evolution leads to a collision with either one of the stars or the gas giant. Finally, we defined the 
dynamical lifetime of particles inside a specific MMR as the time required for 
50\% of the population to leave the 
resonance \citep{gladman97}. In Fig. \ref{F:MMR}, we show the dynamical lifetime in Myr of particles near the internal 
MMRs. 
In the top panel of this figure, the influence of $M_{\scriptscriptstyle 
{\text{b}}}$ is shown for a certain periapsis distance $q_{\scriptscriptstyle 
{\text{b}}}$ = 35 au. The bottom panel summarizes the results for a certain 
mass of the secondary star (i.e. G-type star) and different periapsis 
distances of this star. For the top panel, we have chosen $q_{\scriptscriptstyle 
{\text{b}}}$ = 35 au as for this particular value, the secular resonance 
overlaps with the MMRs in $\mathcal{R}_2$. We can see that prior to the 8:3 MMR, the border between 
icy and rocky asteroids, a secondary M-type will favour chaos inside the rocky bodies region located in 
$\mathcal{R}_1$. This is not surprising since Fig. \ref{F:e_inner} clearly shows the secular resonance overlapping 
with MMRs located inside the snow-line at 2.7 au. Beyond this limit, a higher 
value of $M_{\scriptscriptstyle {\text{b}}}$ leads to a lower dynamical lifetime -- values can reach 0.1 Myr -- since the 
secular resonance  moves outward. From the bottom panel, one can 
recognize that the lower $q_{\scriptscriptstyle 
\text{b}}$, the 
lower the dynamical lifetime. Some MMRs 
can be quickly emptied within 0.1 Myr. With these tests, we highlight that in binary star 
systems, the dynamical 
lifetime of particles initially orbiting inside or close to MMRs can be variable according to the location of the 
secular resonance. Provided that particles can reach the HZ region before colliding with one of the massive bodies 
(i.e. the stars or the gas giant) or before being ejected out of the system, 
they could rapidly cause an early bombardement on any embryos or planets moving in the HZ.

\section{Asteroid flux and water transport to the HZ}\label{S:flux}

\begin{figure*}
\centering{
  \begin{tabular}{c}
  \includegraphics[angle=-90,width=\textwidth]{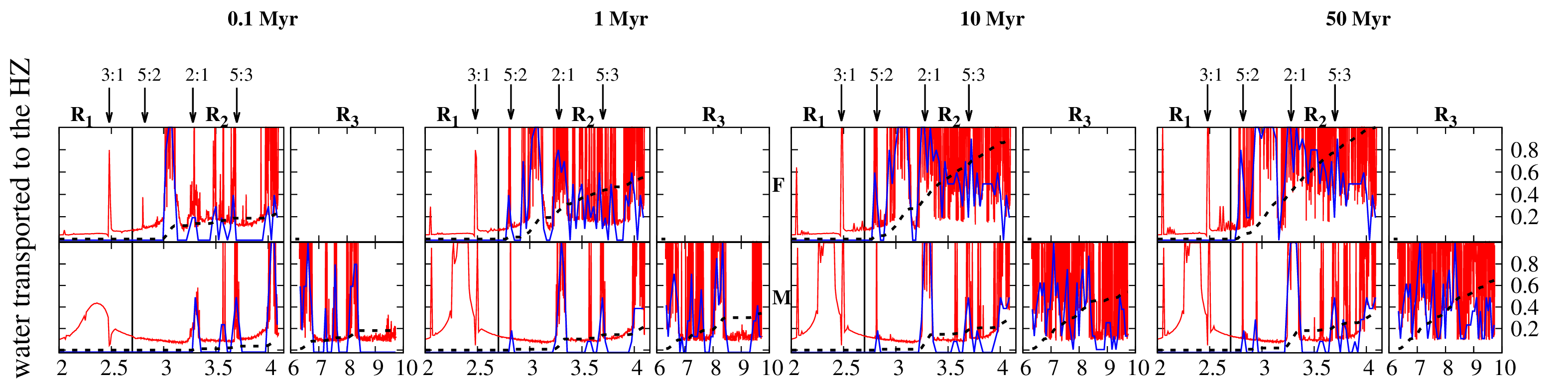}\\
  \includegraphics[angle=-90,width=\textwidth]{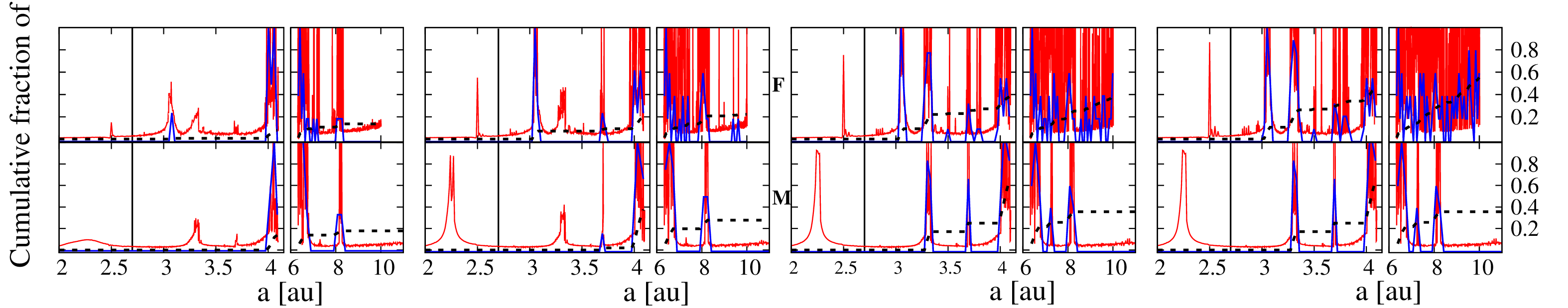}\\
  \end{tabular}}
  \caption{Represented on the y-axis (left and right axes have the same scale), the variation of the 
maximum eccentricity (red line) together with the cumulative fraction of water (dashed black line) brought to the HZ, 
with respect to the initial location of small bodies in $\mathcal{R}_2$ and $\mathcal{R}_3$. The top panel is for a 
secondary star at $q_{\scriptscriptstyle \text{b}}$ = 35 au and the bottom panel is for $q_{\scriptscriptstyle 
\text{b}}$ = 45 au. In this figure, is also represented the normalized HZc distribution (blue line) from 0.1 
Myr (left) up to 50 Myr (right). The vertical black line refers to the position of the snow line. The entire region 
$\mathcal{R}_3$ is intentionally not displayed because of its size. Instead, only the dynamically interesting part of 
$\mathcal{R}_3$ is shown.}
     \label{F:e_HZc}
\end{figure*}

\begin{figure*}
\centering{\includegraphics[angle=-90,width=\textwidth]{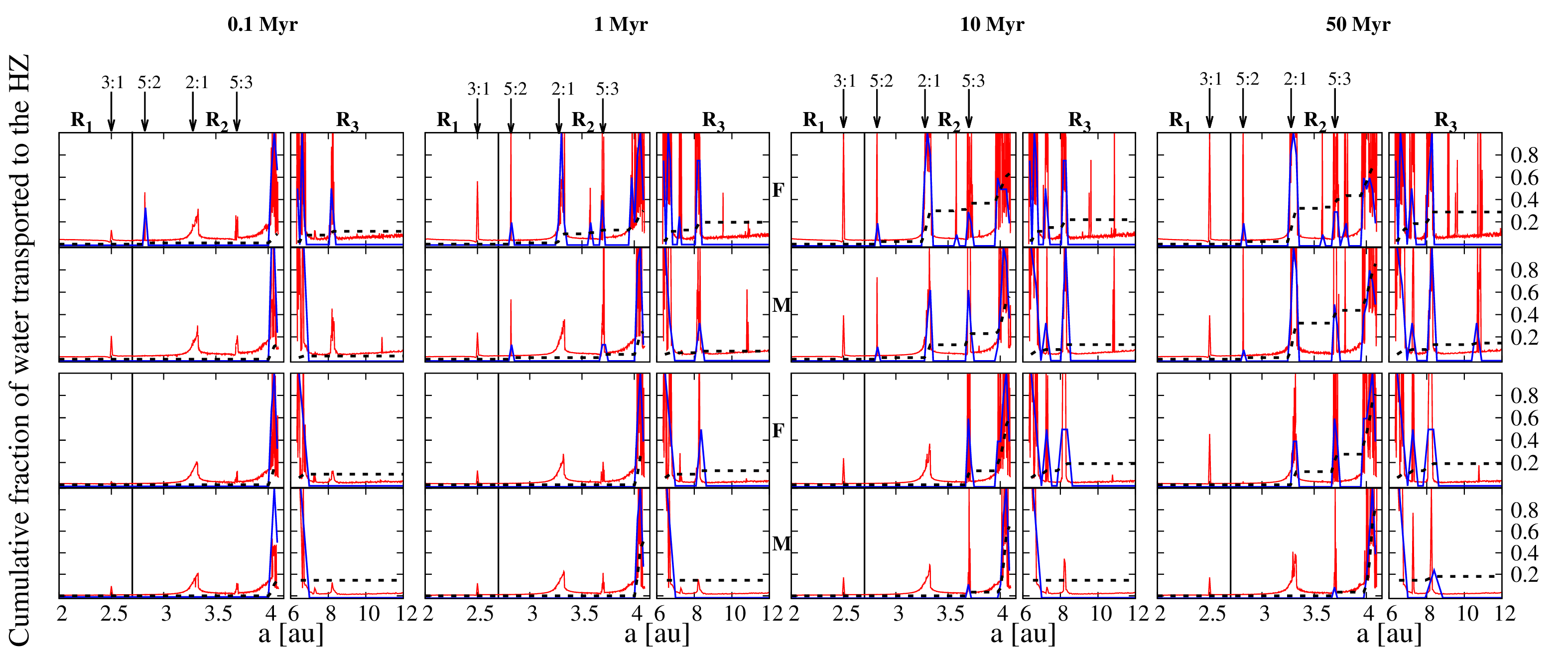}}
  \caption{Same as for Fig. \ref{F:e_HZc} but for $q_{\scriptscriptstyle \text{b}}$ = 70 au (top panel) and 
$q_{\scriptscriptstyle \text{b}}$ = 90 au (bottom panel).}
     \label{F:e_HZc_100}
\end{figure*}

In this section, we compare the flux of icy particles from $\mathcal{R}_2$ and 
$\mathcal{R}_3$ towards the HZ. We recall the definition of habitable zone crossers (HZc) as defined in Paper 
I. HZc are any particles initially evolving beyond the snow line 
(2.7 au) and crossing the HZ 
at least once within the integration time. The authors of Paper I showed that the higher $M_{\scriptscriptstyle 
\text{b}}$  
and $q_{\scriptscriptstyle \text{b}}$, the more important the rate of HZc and, therefore, the quantity of water 
transported to the HZ. We analyze here how the orbital resonances influence the asteroid flux to the HZ and the
amount of water borne therein. On the y-axis in Fig. \ref{F:e_HZc},
we represent the evolution of the maximum 
eccentricity 
(red line) of particles inside $\mathcal{R}_1$, $\mathcal{R}_2$ (as already drawn in Fig. \ref{F:e_inner}) and 
$\mathcal{R}_3$. The top panel corresponds to a secondary star at $q_{\scriptscriptstyle 
\text{b}}$ = 35 au and the bottom panel is for $q_{\scriptscriptstyle 
\text{b}}$ = 45 au, each subpanel corresponding either to 
a secondary F- or M-type star. We chose these two cases to highlight the impact 
of the location of the secular resonance when it lies inside the snow line (M-type) and beyond the 
snow line (F-type). From left to right, 
each figure is for a different period of integration (between 0.1 Myr to 50 
Myr). We also show the normalized HZc distribution (blue line) calculated regarding the total 
number of HZc produced 
by the corresponding systems for each period of integration time. For all  cases considered, within 10 Myr, the 
2:1 MMR located at 
$\sim$ 3.28 au and the secular resonance, when it lies beyond the snow line, are the 
primary sources of HZc in the inner disk.  In addition, the external disk can 
produce a non-negligible or equivalent number of HZc, compared to the inner disk. Indeed, as mentioned in 
Section \ref{S:disk}, the smaller is $q_{\scriptscriptstyle \text{b}}$ the more 
compact is the outer disk. In this case, the depletion can be fast and the 
dynamics are more chaotic. As a result, asteroids might not cross the HZ 
before colliding with the stars and the gas giant or being ejected.
To estimate the amount of transported water, we followed the same 
approach as in Paper I, i.e. the 
contribution of the maximum water content of all the HZc when they first cross 
the HZ. On the same figure, the $y$-axis also corresponds  
to the cumulative fraction of water (dashed black line) brought by the HZc. This fraction is 
determined with respect to the final amount of transported water from $\mathcal{R}_2$ and $\mathcal{R}_3$ 
within 
50 Myr. As already mentioned in Section 
\ref{S:disk}, a system with an F-star as a companion and $q_{\scriptscriptstyle \text{b}}$ = 35 au cannot 
host an outer disk. Therefore, all the incoming water in the HZ for this particular system is from $\mathcal{R}_2$. 
All our systems exhibit the same trend: the quantity of incoming water inside the HZ drastically 
increases when particles orbit initially inside the secular resonance and the 2:1 
MMR. We show similar results
in Fig. \ref{F:e_HZc_100} for $q_{\scriptscriptstyle \text{b}}$ = 70 au (top panel) and $q_{\scriptscriptstyle 
\text{b}}$ = 90 au (bottom panel). 
Contrary to the previous case, the secular resonance does not 
contribute at all to bearing water material into the HZ since it lies in this region 
(see Fig. \ref{F:e_inner}). We show that the two main sources of HZc in $\mathcal{R}_2$ 
are the 2:1 and 5:3 MMR. The contribution of $\mathcal{R}_3$ is more negligible 
than in the previous case since its size is more extended and weakly perturbed. 
\section{Influence of the dynamical parameters on the planetesimal disk}\label{S:parameter}

We are  interested in how the previous results, mainly the origin of HZc and the water transport to the HZ, 
are influenced by the gas giant's dynamical and physical parameters and when, considering the case of 
one or two gas giants that orbit a single star. The same number of particles were placed in the three 
regions\footnote{The size of $\mathcal{R}_{\scriptscriptstyle 2}$ and $\mathcal{R}_{\scriptscriptstyle 3}$ 
can vary because of $R_{\scriptscriptstyle {\text{H},\text{GG}}}$} as defined in Section \ref{S:disk}, but we did not 
repeat the simulations for all the binary systems 
investigated in the previous sections. Instead, we selected two for comparison: a secondary 
F-type with $q_{\scriptscriptstyle \text{b}}$ = 45 au and a secondary M-type with $q_{\scriptscriptstyle \text{b}}$ = 70 
au. Thus, a comparison of the results from this section and illustrated in Fig. \ref{F:e_HZ_all} is 
made in the bottom panels of Fig. \ref{F:e_HZc} (top figure for the F-type) and top panels in Fig. \ref{F:e_HZc_100} 
(bottom figure for the M-type). The results shown in Fig. \ref{F:e_HZ_all} correspond only to 10 Myr of 
integration time.  

\subsection{Gas giant's orbital and physical parameters}\label{S:Jupiter}

\begin{figure*}
\centering{\includegraphics[angle=-90,width=\textwidth]{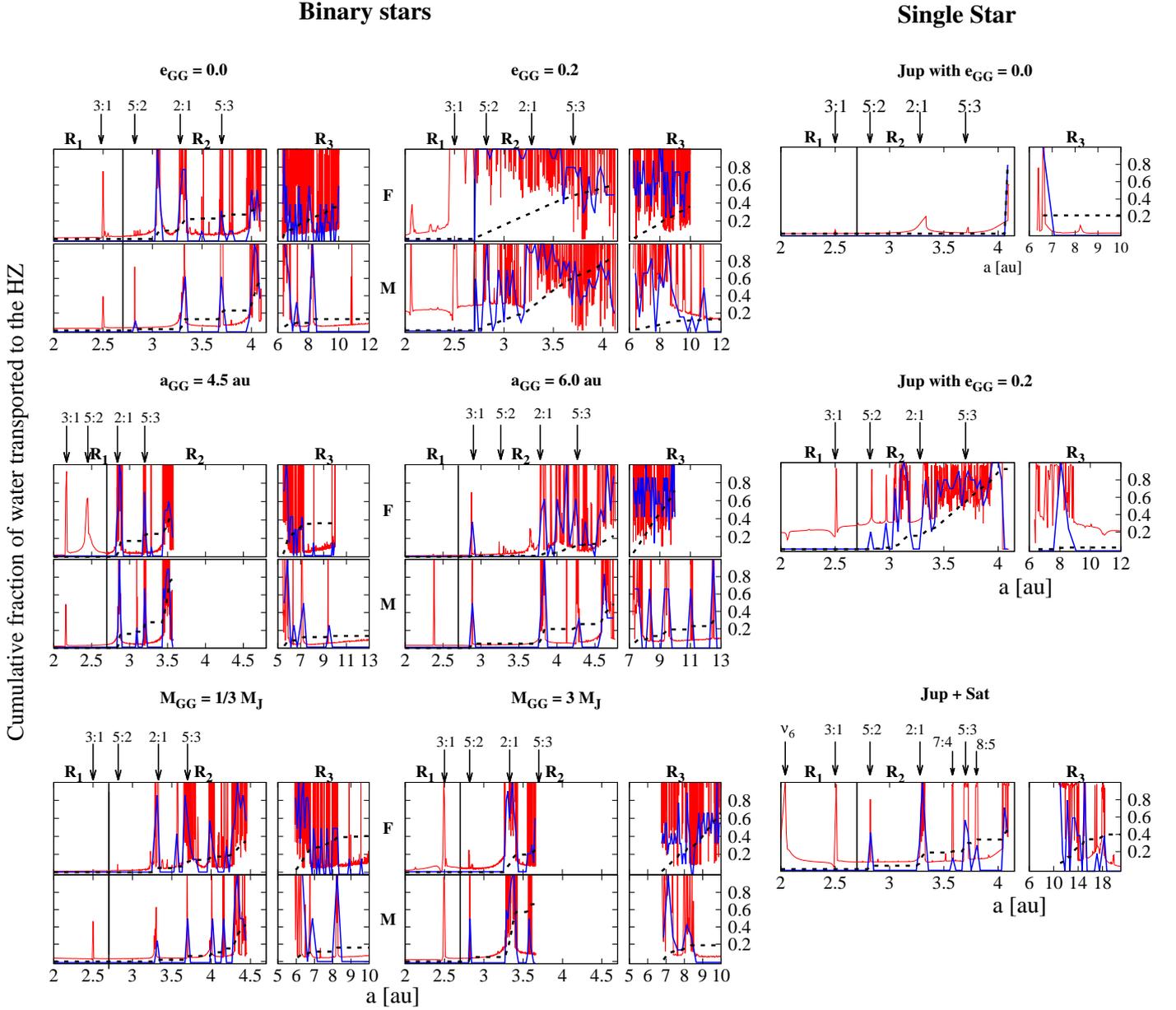}}
  \caption{Left: influence of the gas giant's orbital and physical parameters. Top and bottom subpanels are for 
a secondary F-type and M-type star, respectively. Right: the three subpanels show results for the 
case of single star systems. Each plot corresponds to an intermediate integration time of 10 Myr. See text in Section 
\ref{S:parameter} and legend of Fig. \ref{F:e_HZc} for more details.}
     \label{F:e_HZ_all}
\end{figure*}

We investigated different cases to separately highlight the influence of the initial parameters 
$a_{\scriptscriptstyle \text{GG}}$, $e_{\scriptscriptstyle \text{GG}}$, and $M_{\scriptscriptstyle \text{GG}}$. 
The results are shown on the left panels in Fig. \ref{F:e_HZ_all}: 
\begin{itemize}
 \item Case 1 (top panel): we increase the initial eccentricity to  $e_{\scriptscriptstyle 
\text{GG}}$ = 0.2. For both cases (F- and M-types), the participation of 
$\mathcal{R}_{\scriptscriptstyle 2}$ in the 
water transport to the HZ is much more important than $\mathcal{R}_{\scriptscriptstyle 3}$, since region 
$\mathcal{R}_{\scriptscriptstyle 2}$ is mainly dominated by chaos. Indeed, in addition to stronger interactions with 
the disk, the width of the secular resonance will be increased, for the case of the secondary F-type star, as 
pointed out in \cite{pilat15}. Both strong mechanisms explain the higher flux of asteroids towards the HZ in comparison 
with $e_{\scriptscriptstyle \text{GG}}$ = 0.0.\\

\item Case 2 (middle panel): we change the initial gas giant's semimajor axis to 
$a_{\scriptscriptstyle \text{GG}}$ = 4.5 au and 6.0 au. As a consequence, the secular resonance will be 
shifted inward or outward respectively. For a secondary F-type, for instance, the semi-analytical method developed by 
\cite{pilat15} predicts an inward shift close to the 5:2 MMR (inside the snow-line) or an outward shift beyond the 
2:1 MMR, respectively. Furthermore, in both cases (F- and M-types), decreasing $a_{\scriptscriptstyle 
\text{GG}}$ to 4.5 au  results in shifting the 5:2 MMR inside $\mathcal{R}_1$ (therefore it will not participate in 
bearing water to the HZ), whereas increasing $a_{\scriptscriptstyle \text{GG}}$ to 6.0 au  results in shifting the 3:1 
MMR inside $\mathcal{R}_2$ (thus it  participates in moving icy asteroids from beyond the snow line to the HZ 
region).\\

\item Case 3 (bottom panel): we change the mass of the gas giant to 
$M_{\scriptscriptstyle \text{GG}}$ = 3$M_{\scriptscriptstyle \text{J}}$ 
and $M_{\scriptscriptstyle \text{J}} /\,3$. In the case of a secondary F-type star, for instance, the 
theoretical location of the secular resonance is shifted inward for $M_{\scriptscriptstyle 
\text{GG}}$ = 3$M_{\scriptscriptstyle \text{J}}$ (the secular resonance is below the 3:1 MMR) and outward for 
$M_{\scriptscriptstyle \text{GG}}$ = $M_{\scriptscriptstyle \text{J}} /\,3$. For both secondaries (F- and 
M-types), the MMRs produce 
a significant amount of HZc within 0.1 Myr for $M_{\scriptscriptstyle \text{GG}}$ = 3$M_{\scriptscriptstyle \text{J}}$ 
and 10 Myr for $M_{\scriptscriptstyle \text{GG}}$ = $M_{\scriptscriptstyle \text{J}} /\,3$ whereas 1 Myr is needed for 
$M_{\scriptscriptstyle \text{GG}}$ = 1$\, M_{\scriptscriptstyle \text{J}}$.

\end{itemize}

\subsection{Single star case}\label{S:single}

In case only one star is present in the system, we aim to compare the previous 
results with the case of one or 
two giant planets orbiting a G2V star. We study three possibilities for the giant planets' configurations and 
the corresponding results are shown in the right panels of Fig. \ref{F:e_HZ_all}: 
\begin{itemize}
 \item Case 1 (top panel): one Jupiter at 5.2 au initially on a circular orbit
 \item Case 2 (middle panel): one Jupiter at 5.2 au initially on an elliptic orbit 
with $e_{\scriptscriptstyle \text{GG}}$ = 0.2
 \item Case 3 (bottom panel): a system with two giant planets, i.e. a Jupiter and 
a Saturn\footnote{their mass and initial orbital elements are the commonly used ones}
\end{itemize}

For the three cases, the value of the outer border of 
$\mathcal{R}_3$ is equal to the highest value of $a_{\scriptscriptstyle 
\text{c}}$ in Table \ref{T:ac} (i.e. a wide outer disk). The value of 
the inner border of $\mathcal{R}_3$ is calculated taking into account the Hill radius 
of the outer giant planet (i.e. the Jupiter-like for one gas giant and the 
Saturn-like if the system contains two gas giants). Unsurprisingly, because of the lack of strong 
perturbations\footnote{there cannot be a secular resonance; only MMRs 
can perturb the asteroids' orbits}, only small bodies initially orbiting close to the gas 
giant can 
become HZc (case 1). The flux of icy asteroids drastically increases as soon 
as the initial eccentricity of the gas giant is increased (case 2): the 
interaction with the disk is stronger and it is not easy  to identify the 
most efficient MMR that is producing HZc but the region $\mathcal{R}_2$ is the main 
source of water. In contrast, the contribution of $\mathcal{R}_3$ in  case 
3 is as important  for $\mathcal{R}_2$ because of the presence of the second 
gas giant. Contrary to the binary cases in which the 2:1 MMR and the secular 
resonance (when lying inside $\mathcal{R}_2$, see Fig. \ref{F:e_HZc}) were the 
primary sources of HZc within the whole integration time, in case 3, the 2:1 MMR becomes 
dominant only within 10 Myr of integration. Below 10 Myr, the 5:2 and 5:3 MMRs 
dominate over the other MMRs. 

\begin{figure*}
\centering{
\begin{tabular}{c}
\includegraphics[angle=-90,width=0.9\textwidth]{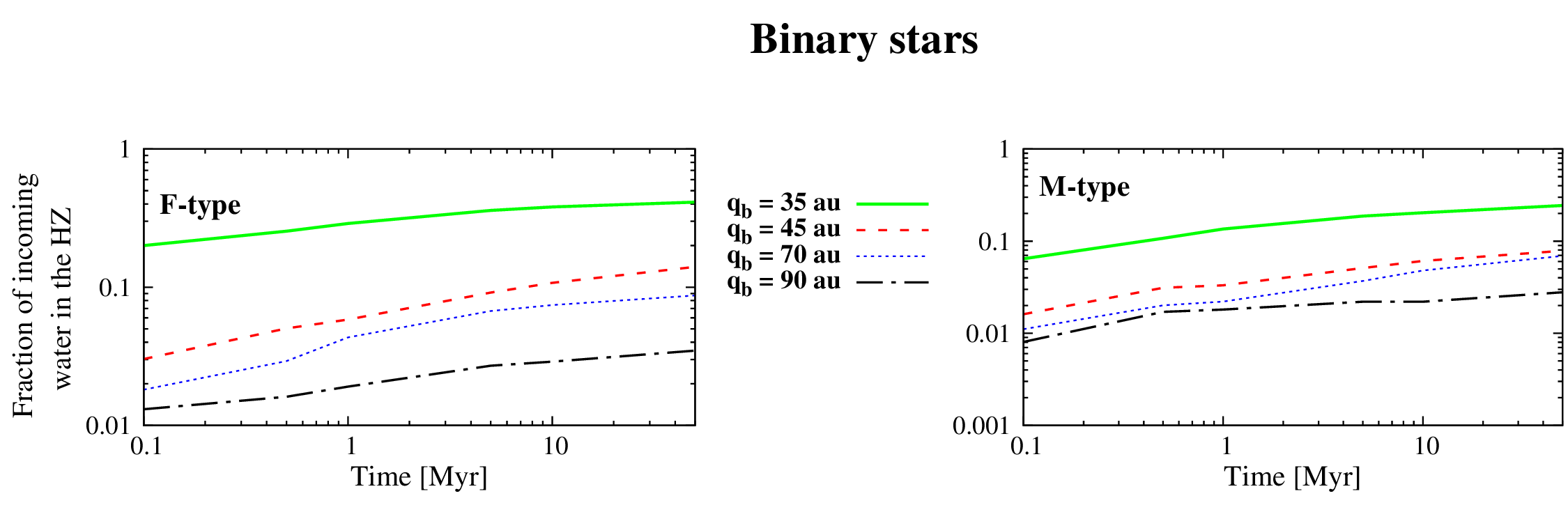}\\
\includegraphics[angle=-90,width=0.9\textwidth]{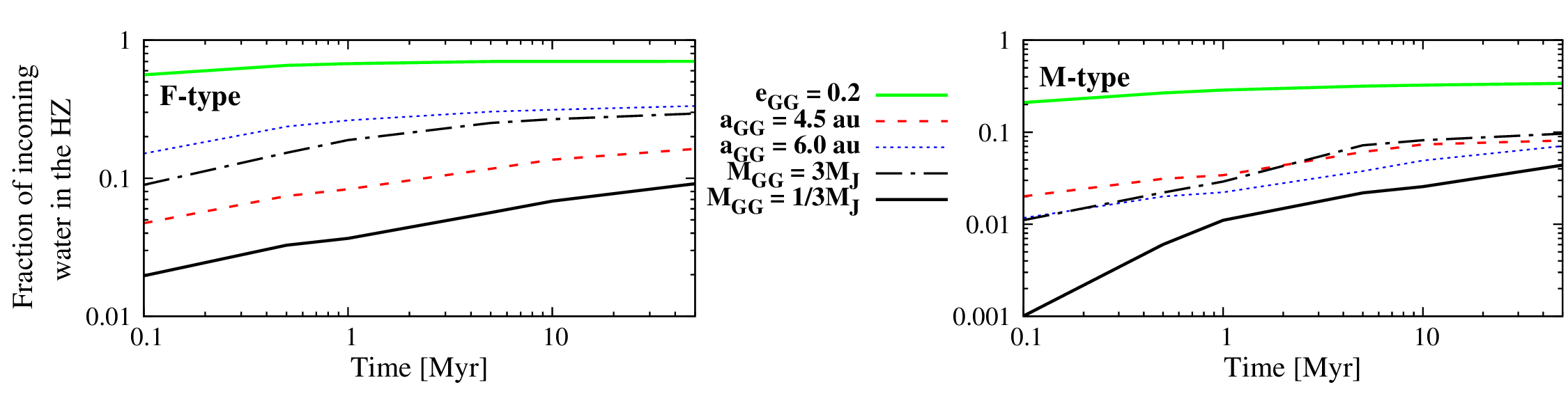}\\
\includegraphics[angle=-90,width=0.5\textwidth]{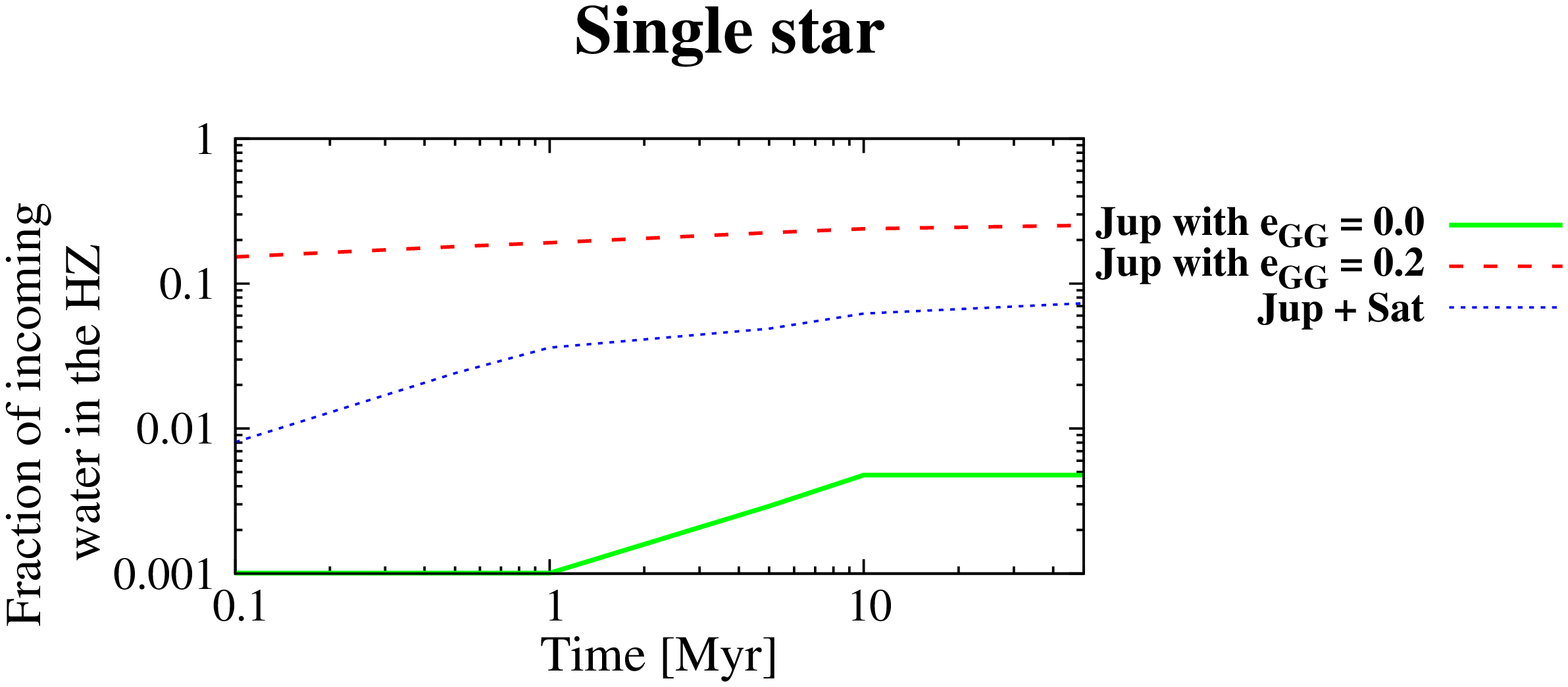}\\
\end{tabular}}
  \caption{Fraction of incoming water in the HZ with respect to the initial total amount of water in the disk 
of planetesimals, considering various configurations. Top and middle panels show results in binary star systems for 
fixed and variable values of the gas giant's orbital and physical parameters, respectively, when the secondary is an 
F-type (left) and M-type (right) star (see text in Section \ref{S:Jupiter}). Comparison with various single star 
systems are shown in the bottom panel (see text in Section \ref{S:single}).}
     \label{F:water_article}
\end{figure*}

\section{Comparison of the water transport efficiency}\label{S:comparison}

Finally, in Fig. \ref{F:water_article}, we combine the results obtained in 
Sections \ref{S:flux} and \ref{S:parameter} to compare 
the total amount of water transported into the HZ, expressed with respect to 
the initial total amount of water in the disk. The top panel compiles the 
results obtained in Section \ref{S:flux} for a secondary F-type (left) and 
M-type (right). The middle panel compiles the results from Section 
\ref{S:Jupiter} when changing the gas giant's orbital and physical parameters, 
for a secondary F-type at $q_{\scriptscriptstyle \text{b}}$ = 45 au and a 
secondary M-type at $q_{\scriptscriptstyle \text{b}}$ = 70 au. Finally, the 
bottom panel shows the results for single star systems containing one or two 
gas giants.\\
In the binary cases, when the parameters of the gas giant are fixed (top panels), one can see for instance, that a 
secondary M-type star at $q_{\scriptscriptstyle \text{b}}$ = 35 au, needs $\sim$50 Myr so that one quarter of the 
water initially in the disk of planetesimals is transported to the HZ, contrary to an F-type star where only $\sim$0.5 
Myr is needed. This is mainly due to 
the presence of the secular resonance inside the asteroid belt in $\mathcal{R}_2$. \\
When changing the gas giant's orbital and physical parameters (middle panels), the water transport efficiency 
to the HZ can be different (comparisons have to be made with the red-dashed line for the secondary F-star at 
$q_{\scriptscriptstyle \text{b}}$ = 45 au and the blue-dotted line for the M-type at $q_{\scriptscriptstyle \text{b}}$ = 
70 au in the top panels). For instance, the figure reveals that  increasing $e_{\scriptscriptstyle 
\text{GG}}$ boosts the water 
transport efficiency. Indeed, within 0.1 Myr, almost 50\% of the initial water ended up in the HZ in the case of an 
F-star, and nearly 20\% for an M-type\footnote{For both cases, this is 2.5 times more than for the case  
$e_{\scriptscriptstyle \text{GG}}$ = 0 }. When 
changing $a_{\scriptscriptstyle \text{GG}}$, even with the lack of active orbital resonances in the case of 
$a_{\scriptscriptstyle \text{GG}}$ = 4.5 au (the 5:2 MMR and the secular resonance lie inside the snow-line), the 2:1 
and 5:3 MMR are powerful enough so that the water transport is as efficient as for a giant 
planet at $a_{\scriptscriptstyle \text{GG}}$ = 5.2 au. For $a_{\scriptscriptstyle \text{GG}}$ = 6.0 au, since 
the 3:1 MMR participates in the asteroid flux, the water transport is even more efficient. The mass of the gas giant 
also influences the final total amount of water brought to the HZ as a higher or smaller value of 
$M_{\scriptscriptstyle \text{GG}}$,  respectively, strengthens or weakens the water transport efficiency of a binary 
star system. \\
Finally, comparable results (bottom panel) can be 
obtained with different single star configurations hosting either one or two 
giants planets. If only one giant planet, with $e_{\scriptscriptstyle \text{GG}}$ = 0.0, 
orbits a sun-like star, a significant water transport in the HZ within 50 Myr is very unlikely because of the lack of 
gravitational perturbations. However, if the gas giant initially starts on an eccentric orbit ($e_{\scriptscriptstyle 
\text{GG}}$ = 0.2), the depletion of $\mathcal{R}_2$  is quite fast and within 0.1 Myr, the water transport 
efficiency can be comparable to a secondary F-type star at $q_{\scriptscriptstyle \text{GG}}$ = 45 au within 50 Myr. 
This is not surprising since such a high eccentricity  favours chaos in $\mathcal{R}_2$. Last but not least, if two giant 
planets (Jupiter- and Saturn-like) orbit a sun-like star, both regions $\mathcal{R}_2$ and $\mathcal{R}_3$ are water 
sources since several inner and outer MMRs with the giant planets are active and 50 Myr are needed to transport 10\% of 
the 
water initially present in the disk. These results are comparable to the binary star systems simulations with an M-type 
as a companion.

\section{Conclusion}\label{S:conclusion}

We showed that the flux of icy bodies towards circumprimary HZs in 
binary star systems can vary 
according to the characteristics and motion of the secondary star and the giant planet. First, we 
showed that a gas giant planet can suffer from both variations of its 
orbital eccentricity and a drift in semi-major axis. This in turn could 
strengthen the interaction with an inner (region $\mathcal{R}_2$) and outer 
(region $\mathcal{R}_3$) disk of planetesimals. 
Moreover we highlight that in tight binaries ($a_{\scriptscriptstyle \text{b}}$ = 50 au), a secular resonance can 
lie within the inner asteroid belt, 
overlapping with MMRs, which enable, in a short timescale, an efficient and 
significant flux of icy asteroids towards the 
HZ, in which particles orbit in a near circular motion (region $\mathcal{R}_1$). 
By way of contrast, in the study of wide binaries 
($a_{\scriptscriptstyle 
\text{b}}$ = 100 au), particles inside the HZ can 
move on eccentric orbits when the secular resonance lies in the HZ. The 
outer asteroid belt is only perturbed by MMRs. As a 
consequence, a longer timescale is needed to produce a significant flux of icy 
asteroids towards the HZ. These dynamics  
drastically impact the dynamical lifetime of particles initially located inside inner MMRs. Indeed, this can range from 
thousands of years to several million years according to the location of the MMR within the secular resonance. This 
can favour a fast and significant contribution of MMRs in producing HZc, which are asteroids with orbits crossing the HZ and 
bearing water therein. In any case, we highlighted that, for the studied binary 
star systems, the inner disk (region $\mathcal{R}_2$) is the primary source of 
HZc (and therefore water in the HZ), by  means of the 
2:1 MMR, the 5:3 MMR, and the secular resonance, when this latter lies close 
or beyond the snow line. As shown in Section \ref{S:Jupiter}, the gas giant's characteristics also influence 
the asteroids flux to the HZ and therefore the water transport. Indeed, the dynamical interactions can be different since 
the location of the orbital resonances can be shifted inward or outward. Giant planets, initially on eccentric orbits, 
are an efficient way to ensure that the HZ can be rapidly fed with water (within 0.1 Myr) since it can increase the width 
of the secular resonance for instance. Even gas giants with lower mass can be efficient in the water transport to the 
HZ but on a longer timescale. For any 
giant planet configuration studied, the 2:1 MMR and 5:3 MMR also appeared to be powerful perturbations for transporting
water into the HZ. This is not necessarily the case in single star systems. In Jupiter- and Saturn-like systems 
orbiting a sun-like star, other inner and outer MMRs are also, to a lesser extent, water sources. \\

As for the amount of transported water that effectively ends up on embryos and planets in the HZ, both Fig. 
\ref{F:e_inner} and Paper I have pointed out two opposite behaviours that depend on the binary's 
orbit: a high rate of HZc and 
nearly circular motion in the HZ (tight binary) versus low rate of HZc and eccentric orbits in the HZ (wide binary). The 
main problems of the tight-binary case are: 
\begin{itemize}
 \item [a)] planets or embryos moving on nearly circular motion in the HZ  have lower impact probabilities 
with HZc -- with regard also to the water distribution within the HZ as shown in Paper I -- than if they were 
moving 
on 
eccentric orbits;
\item [b)] the asteroid flux to the HZ is a fast process. If the primary star is in the T-Tauri phase, 
\cite{tu15} show that the activity of a present Sun-like star can have a different history because of its 
rotational behaviour. As a consequence, planetary embryos might not be able to keep the incoming water on 
their surface;
\item [c)] with respect to this last point, and that the depletion of the inner disk 
(region $\mathcal{R}_2$) can be quite fast, there would not be 
other 
water sources available after the activity of the primary star has significantly decreased, if the remaining 
asteroids 
(in the inner and outer disk) stay on stable orbits.
\end{itemize}
In wide binaries, since the flux process is much slower, the activity of a young primary star would not be an obstacle 
for a planet to keep the water borne from asteroids on its surface, even if it loses its primary surface 
water 
content. Indeed, the eccentric motion inside the HZ can have a positive aspect: since planetary embryos can 
leave the HZ 
from time to time, they can increase their impact probabilities since they can also 
interact with icy asteroids evolving 
beyond the outer border of the HZ. However, the high eccentric motion inside the HZ can also raise 
some problems that would be subject to future study.

\begin{acknowledgements}

DB and EPL acknowledge the support of the Austrian Science Fundation (FWF) NFN project: Pathways to Habitability, 
subproject
S11608-N16 Binary Star Systems and Habitability. DB and EPL acknowledge also the Vienna
Scientific Cluster (VSC project 70320) for computational resources. AB and EPL acknowledge 
the support of the FWF project P22603-N16.

\end{acknowledgements}

\bibliographystyle{plainnat}

\bibliography{biblio} 

\end{document}